\documentclass[twoside,12pt]{article}
\usepackage{CJK}
\usepackage{indentfirst}
\usepackage{bm}
\usepackage{verbatim}
\usepackage{graphicx}
\usepackage{multicol}

\begin{document}
\begin{CJK*}{GBK}{song}
\newcommand{\song}{\CJKfamily{song}}
\newcommand{\hei}{\CJKfamily{hei}}
\newcommand{\fs}{\CJKfamily{fs}}
\newcommand{\kai}{\CJKfamily{kai}}
\def\thefootnote{\fnsymbol{footnote}}
\begin{center}
\Large\hei  Analytical solution and entanglement swapping of a double Jaynes-Cummings model in non-Markovian environments 
\end{center}

\footnotetext{\hspace*{-.45cm}\footnotesize zhmzc1997@126.com, tel:13807314064}

\begin{center}
\rm Hong-Mei Zou, \ Mao-Fa Fang
\end{center}

\begin{center}
\begin{footnotesize} \rm
Key Laboratory of Low-dimensional Quantum Structures and Quantum Control of Ministry of Education, College of Physics and Information Science, Hunan Normal University, Changsha, 410081, China\  \\   

\end{footnotesize}
\end{center}

\begin{center}
\footnotesize (Received XX; revised manuscript received X XX)
\end{center}

\vspace*{2mm}

\begin{center}
\begin{minipage}{15.5cm}
\parindent 20pt\footnotesize

Analytical solution and entanglement swapping of a double Jaynes-Cummings model in non-Markovian environments are investigated by the timeconvolutionless master equation method. We obtain the analytical solution of this model and discuss in detail the influence of atom-cavity coupling, non-Markovian effect and initial state purity on entanglement dynamics. The results show that, in the non-Markovian environments, the entanglement between two cavities can be swapped to other bipartite subsystems by interaction between an atom and its own cavity. Due to the dissipation of environment, the entanglements of all bipartite subsystems will eventually decay to zero when the atom couples weakly to its cavity and the non-Markovian effect is also weak. All bipartite subsystems can tend to steady entanglement states if and only if there is the strong atom-cavity coupling or the strong non-Markovian effect. The steady state of the subsystem composed of an atom and its own cavity is independent on the purity but the steady states of other bipartite subsystems are dependent on the purity.

\end{minipage}
\end{center}

\begin{center}
\begin{minipage}{15.5cm}
\begin{minipage}[t]{2.3cm}{\hei Keywords: }\end{minipage}
\begin{minipage}[t]{13.1cm}analytical solution, entanglement swapping, double Jaynes-Cummings model, non-Markovian environment
\end{minipage}\par\vglue8pt
{\bf PACS: }03.65.Yz, 03.65.Ud, 03.67.Bg, 42.50.Pq
\end{minipage}
\end{center}

\section{\fs Introduction}  
Entanglement has been considered as one of the most important resources for quantum information processing and quantum communication including quantum key distribution and quantum secret sharing\cite{Schrodinger,Nielsen}. Therefore, a great deal of attention has been devoted to the experimental generation and manipulation of entangled systems and the theoretical study of entanglement evolution\cite{Raimond,Julsgaard,Blinov,Haroche}. In particular, since Yu and Eberly\cite{Eberly} discovered that the Markovian entanglement dynamics of two qubits exposed to local noisy environments may markedly differ from a single qubit decoherence evolution, it has become an important topic that the analysis of entanglement decay and its relation with decoherence induced by unavoidable interaction between a system and its environment. Numerous investigations on the entanglement dynamics in noisy environments have been done for more than a decade. These studies may be roughly separated into three categories. The first one is that presence of environment noise makes the quantum entanglement of two qubits decay exponentially and vanish asymptotically \cite{bellomo3,breuer4,guohong1,anjh1,anjh4,Xiao1,CaiJ}. The second is that the environmental noise can make entanglement sudden death and entanglement sudden birth\cite{mazzola1,zhengh1,xuzy,Wolf}. In the third category, the environmental noise can be helpful to keeping the system entangled in the steady state\cite{guohong1,Altintas,Memarzadeh}. More important, in the last year, the authors in\cite{Fanchini,Haseli} investigated the information exchange between an open quantum system and its environment by using the quantum loss. They found that the entanglement-based measure of non-Markovianity is related to the flow of information between the quantum system and its environment and had presented an experimental realization of this scenario. The authors in\cite{Reina} studied the qubit-environment dynamics from a different perspective by means of the Koashi-Winter relation\cite{Fanchini2,Koashi} and showed that valuable information about the evolution of quantum entanglement and correlations can be obtained if the flow of information between the register and the environment is better understood.

It is well known that the Jaynes-Cummings model(JCM) is the simplest possible physical model that describes the interaction of a two-level atom with a single-mode cavity\cite{Jaynes}, and has been used to understand a wide variety of phenomena in quantum optics and condensed matter systems, such as trapped ions, quantum dots, superconducting circuits, optical and microwave cavity QED, among others\cite{Law,Strauch1,Childs,Ben-Kish,Strauch2,Brion}. Afterwards, T. Yu \emph{et al} proposed a double Jaynes-Cummings model(DJCM)\cite{Yonac1,Yonac2} which consists of two separate JCM systems. Recently, the DJCM systems have been extensively investigated\cite{Sainz,Vieira,Veitia}.

Although many important progresses have been acquired in experimental and theoretical researches on the entanglement dynamics in quantum systems, these investigations mentioned above are mainly focused on the two classes of model. The one is the model of open quantum systems of bipartite directly interacting with a non-Markovian environment or two non-Markovian environments. Another is the model of closed quantum systems of bipartite interacting with a cavity or two cavities, in which the influence of environment on the qubit-cavity system is neglected. In fact, any qubit-cavity systems are all open so that they will inevitably interact with external environment. Thus, how to solve the open qubit-cavity systems becomes a very important topic. But, at present, the research on two qubit-cavity system in non-Markovian environment has not been yet reported. Here, we approach the DJCM in non-Markovian environments by means of the timeconvolutionless(TCL) master equation method\cite{Breuer}, in which the atom $A$ interacts only with the cavity $a$, where the cavity is coupled to the bosonic environment $1$, and similar for the atom $B$ and the cavity $b$ as well as the environment $2$. We assume that, the two-cavity initial state is an extension of the Werner state and the two-atom initial state is $|gg\rangle$. This "extended" Wernerlike state is a mixed state that may reduce to a Bell-like pure state\cite{Bellomo1} or to a Werner mixed state\cite{WeiTC}. We derive an analytical solution of this DJCM, which is the first aim of this paper. And we also obtain steady entanglement states of six bipartite subsystems in this DJCM, which is the second goal of this paper.

On the other hand, entanglement swapping, as a fascinating feature of entanglement\cite{Zukowski,Zeilinger,Bose,PanJW}, is the basis of quantum repeaters\cite{Wang}, which allow to distribute entanglement at long distances in an efficient manner. Thus entanglement swapping is recognized as a fundamental tool for quantum communication. Its importance has been also reflected in quantum cryptography\cite{Pirandola}. Recently, it has been demonstrated that the entanglement can be swapped to timelike separated quantum systems\cite{Megidish}. In this paper, we propose a scheme to realize entanglement swapping in non-Markovian environments by means of the atom-cavity coupling, which is the third purpose of this paper.

The outline of the paper is the following. In Section 2, we give an analytical solution of the DJCM in non-Markovian environments. In Section 3, we provide steady entanglement states of bipartite subsystems of the DJCM in non-Markovian environments. In Section 4, we discuss the entanglement swapping of the DJCM in non-Markovian environment. Finally, we conclude with a brief summary of important results in Section 5.

\section{\fs Analytical solution of the DJCM in non-Markovian environments}
We consider a composite system of two two-level atoms($A$,$B$) interacting respectively with two cavities($a$,$b$), where each cavity is coupled to a bosonic environment($1$,$2$), called the DJCM in non-Markovian environments, and there is no interaction between the partition $"Aa1"$ and $"Bb2"$, shown as in Fig.1.

\begin{center}
\includegraphics[width=10cm,height=5cm]{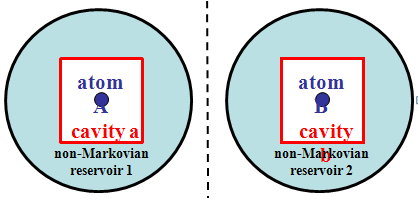}
\parbox{16cm}{\small{\bf Fig1.}
(Color online)A schematic figure of the DJCM in non-Markovian environments. In the left(right) partition there is the atom $A$($B$) interacting with the cavity $a$($b$), respectively, where the cavity is coupled to a bosonic environment $1$($2$), and there is no interaction between the partition $"Aa1"$ and $"Bb2"$.}
\end{center}

 The Hamiltonian of the left partition $"Aa1"$ in Fig.1 is $H=H_{JC}+H_{r}+H_{I}$. At resonance, under the rotating wave approximation\cite{Vieira,Scala1,Scala2}, in units of $\hbar$, $H_{JC}=\frac{1}{2}\omega_{0}\sigma_{z}+\omega_{0}a^{\dag}a +\Omega(a\sigma_{+}+a^{\dag}\sigma_{-})$, $H_{r}=\sum_{k}\omega_{k}c_{k}^{\dag}c_{k}$ and $H_{I}=(a^{\dag}+a)\sum_{k}g_{k}(c_{k}^{\dag}+c_{k})$, where $a^{\dag}$ and $a$ are the creation and annihilation operators of the cavity $a$, $\sigma_{+}=|e\rangle\langle g|$, $\sigma_{-}=|g\rangle\langle e|$, and $\sigma_{z}=|e\rangle\langle e|-|g\rangle\langle g|$, here $|g\rangle$ and $|e\rangle$ denote the atomic ground and excited state, respectively\cite{Jaynes}, $\omega_{0}$ is the atomic Bohr frequency and $\Omega$ is the coupling constant between the atom and its cavity, and $c_{k}^{\dag}$ and $c_{k}$ are the creation and annihilation operators of the reservoir, $g_{k}$ is the coupling constant between the cavity and its reservoir. So the Hamiltonian may be written as
\begin{eqnarray}\label{EB01}
H&=&\frac{1}{2}\omega_{0}\sigma_{z}+\omega_{0}a^{\dag}a+\Omega(a\sigma_{+}+a^{\dag}\sigma_{-})+\sum_{k}\omega_{k}c_{k}^{\dag}c_{k}+(a^{\dag}+a)\sum_{k}g_{k}(c_{k}^{\dag}+c_{k})
\end{eqnarray}

Using the second order of the TCL expansion\cite{Breuer}, neglecting the atomic spontaneous emission and the Lamb shifts, and assuming one initial excitation and a reservoir at zero temperature, the non-Markovian master equation for the density operator $R(t)$ in the dressed-state basis $\{|E_{1+}\rangle, |E_{1-}\rangle, |E_{0}\rangle\}$ is
\begin{eqnarray}\label{EB05}
\dot{R}(t)&=&-i[H_{JC},R(t)]\nonumber\\
&+&\gamma(\omega_{0}+\Omega,t)(\frac{1}{2}|E_{0}\rangle\langle E_{1+}|R(t)|E_{1+}\rangle\langle E_{0}|-\frac{1}{4}\{|E_{1+}\rangle\langle E_{1+}|,R(t)\})\nonumber\\
&+&\gamma(\omega_{0}-\Omega,t)(\frac{1}{2}|E_{0}\rangle\langle E_{1-}|R(t)|E_{1-}\rangle\langle E_{0}|-\frac{1}{4}\{|E_{1-}\rangle\langle E_{1-}|,R(t)\}),
\end{eqnarray}
where $|E_{1\pm}\rangle=(|1g\rangle\pm|0e\rangle)/\sqrt{2}$ are the eigenstates of $H_{JC}$ with one total excitation, with energy $\omega_{0}/2\pm\Omega$, and $|E_{0}\rangle=|0g\rangle$ is the ground state, with energy $-\omega_{0}/2$. The timedependent decay rates for $|E_{1-}\rangle$ and $|E_{1+}\rangle$ are $\gamma(\omega_{0}-\Omega,t)$ and $\gamma(\omega_{0}+\Omega,t)$ respectively.

If the reservoir at zero temperature is modeled with a Lorentzian  spectral density\cite{Scala1}
\begin{equation}\label{EB06}
J(\omega)=\frac{1}{2\pi}\frac{\gamma_{0}\lambda^{2}}{(\omega_{1}-\omega)^{2}+\lambda^{2}},
\end{equation}
where the parameter $\lambda$ defines the spectral width of the coupling, which is connected to the reservoir correlation time $\tau_{R}$ by $\tau_{R}$=$\lambda^{-1}$ and the parameter $\gamma_{0}$ is related to the relaxation time scale $\tau_{S}$ by $\tau_{S}$=$\gamma_{0}^{-1}$. In the subsequent analysis of dynamical evolution of the system, typically a weak and a strong coupling regimes can be distinguished. For a weak regime we mean the case $\lambda>2\gamma_{0}$, that is, $\tau_{S}>2\tau_{R}$. In this regime the relaxation time is greater than the reservoir correlation time and the behavior of dynamical evolution of the system is essentially a Markovian exponential decay controlled by $\gamma_{0}$. In the strong coupling regime, that is, for $\lambda<2\gamma_{0}$, or $\tau_{S}<2\tau_{R}$, the reservoir correlation time is greater than or of the same order as the relaxation time and non-Markovian effects become relevant\cite{Breuer,Bellomo1,Bellomo2}. Supposing the spectrum is peaked on the frequencies of the states $|E_{1\pm}\rangle$, i.e. $\omega_{1}=\omega_{0}\pm\Omega$, the decay rates for the two dressed states $|E_{1\pm}\rangle$ are respectively expressed as\cite{Scala1} $\gamma(\omega_{0}-\Omega,t)=\gamma_{0}(1-e^{-\lambda t})$ and $\gamma(\omega_{0}+\Omega,t)=\frac{\gamma_{0}\lambda^{2}}{4\Omega^{2}+\lambda^{2}}\{1+[\frac{2\Omega}{\lambda}sin2\Omega t-cos2\Omega t]e^{-\lambda t}\}$.

If the system starts from the state
\begin{eqnarray}\label{EB09}
R(0)=\left(
            \begin{array}{cccc}
              R_{11}(0)&R_{12}(0)&R_{13}(0)\\
              R_{21}(0)&R_{22}(0)&R_{23}(0)\\
              R_{31}(0)&R_{32}(0)&R_{33}(0)\\
            \end{array}
          \right),
\end{eqnarray}
we can acquire the matrix elements at all times from Eq.~(\ref{EB05})
\begin{eqnarray}\label{EB10}
      R_{11}(t)&=&A_{11}^{11}R_{11}(0),      R_{12}(t)=A_{12}^{12}R_{12}(0),      R_{13}(t)=A_{13}^{13}R_{13}(0),\nonumber\\
      R_{22}(t)&=&A_{22}^{22}R_{22}(0),      R_{23}(t)=A_{23}^{23}R_{23}(0),\nonumber\\
      R_{33}(t)&=&A_{33}^{11}R_{11}(0)+A_{33}^{22}R_{22}(0)+A_{33}^{33}R_{33}(0),
\end{eqnarray}
here
\begin{eqnarray}\label{EB11}
      A_{11}^{11}&=&e^{-\frac{1}{2}I_{+}},   A_{12}^{12}=e^{-2i\Omega t}e^{-\frac{1}{4}(I_{+}+I_{-})}, A_{13}^{13}=e^{-i(\omega_{0}+\Omega)t}e^{-\frac{1}{4}I_{+}},\nonumber\\
      A_{22}^{22}&=&e^{-\frac{1}{2}I_{-}},   A_{23}^{23}=e^{-i(\omega_{0}-\Omega)t}e^{-\frac{1}{4}I_{-}},\nonumber\\
      A_{33}^{11}&=&1-A_{11}^{11},   A_{33}^{22}=1-A_{22}^{22},      A_{33}^{33}=1,
\end{eqnarray}
and
\begin{eqnarray}\label{EB12}
     I_{-}&=&\gamma_{0}t+\frac{\gamma_{0}}{\lambda}(e^{-\lambda t}-1),\nonumber\\
     I_{+}&=&\frac{\gamma_{0}\lambda^{2}}{4\Omega^{2}+\lambda^{2}}[t-\frac{4\Omega e^{-\lambda t} sin(2\Omega t)}{4\Omega^{2}+\lambda^{2}}+\frac{(\lambda^{2}-4\Omega^{2})(e^{-\lambda t}cos(2\Omega t)-1)}{\lambda(4\Omega^{2}+\lambda^{2})}].
\end{eqnarray}

Next, we extend the JCM in non-Markovian environment to the DJCM in non-Markovian environments. Following the procedure described in Ref.\cite{Bellomo1,Bellomo2}, we construct the density matrix $R^{T}(t)$ for this DJCM in non-Markovian environment. This two JCM subsystems, labeled by $A$ and $B$, respectively, may be in general in different environments so that their evolutions are characterized by the different functions $A_{ij}^{mn}$ and $B_{i'j'}^{m'n'}$. In the dressed-state basis $ \mathcal{D}^{T}=\{|1\rangle\equiv|E_{1+}E_{1+}\rangle, |2\rangle\equiv|E_{1+}E_{1-}\rangle, |3\rangle\equiv|E_{1+}E_{0}\rangle, |4\rangle\equiv|E_{1-}E_{1+}\rangle, |5\rangle\equiv|E_{1-}E_{1-}\rangle, |6\rangle\equiv|E_{1-}E_{0}\rangle, |7\rangle\equiv|E_{0}E_{1+}\rangle, |8\rangle\equiv|E_{0}E_{1-}\rangle, |9\rangle\equiv|E_{0}E_{0}\rangle\}$, using Eq.~(\ref{EB10}) and Eq.~(\ref{EB11}), we get the diagonal elements
\begin{eqnarray}\label{EB13}
      R_{11}^{T}(t)&=&A_{11}^{11}B_{11}^{11}R_{11}^{T}(0),R_{22}^{T}(t)=A_{11}^{11}B_{22}^{22}R_{22}^{T}(0),\nonumber\\      R_{33}^{T}(t)&=&A_{11}^{11}(B_{33}^{11}R_{11}^{T}(0)+B_{33}^{22}R_{22}^{T}(0)+B_{33}^{33}R_{33}^{T}(0)),\nonumber\\
      R_{44}^{T}(t)&=&A_{22}^{22}B_{11}^{11}R_{44}^{T}(0),R_{55}^{T}(t)=A_{22}^{22}B_{22}^{22}R_{55}^{T}(0),\nonumber\\      R_{66}^{T}(t)&=&A_{22}^{22}(B_{33}^{11}R_{44}^{T}(0)+B_{33}^{22}R_{55}^{T}(0)+B_{33}^{33}R_{66}^{T}(0)),\nonumber\\
      R_{77}^{T}(t)&=&B_{11}^{11}(A_{33}^{11}R_{11}^{T}(0)+A_{33}^{22}R_{44}^{T}(0)+A_{33}^{33}R_{77}^{T}(0)),\nonumber\\
      R_{88}^{T}(t)&=&B_{22}^{22}(A_{33}^{11}R_{22}^{T}(0)+A_{33}^{22}R_{55}^{T}(0)+A_{33}^{33}R_{88}^{T}(0)),\nonumber\\
      R_{99}^{T}(t)&=&A_{33}^{11}(B_{33}^{11}R_{11}^{T}(0)+B_{33}^{22}R_{22}^{T}(0)+B_{33}^{33}R_{33}^{T}(0)),\nonumber\\
      &+&A_{33}^{22}(B_{33}^{11}R_{44}^{T}(0)+B_{33}^{22}R_{55}^{T}(0)+B_{33}^{33}R_{66}^{T}(0)),\nonumber\\
      &+&A_{33}^{33}(B_{33}^{11}R_{77}^{T}(0)+B_{33}^{22}R_{88}^{T}(0)+B_{33}^{33}R_{99}^{T}(0)),
\end{eqnarray}
and the nondiagonal elements
\begin{eqnarray}\label{EB14}
      R_{12}^{T}(t)&=&A_{11}^{11}B_{12}^{12}R_{12}^{T}(0),R_{13}^{T}(t)=A_{11}^{11}B_{13}^{13}R_{13}^{T}(0),\nonumber\\
      R_{14}^{T}(t)&=&A_{12}^{12}B_{11}^{11}R_{14}^{T}(0),R_{15}^{T}(t)=A_{12}^{12}B_{12}^{12}R_{15}^{T}(0),\nonumber\\
      R_{16}^{T}(t)&=&A_{12}^{12}B_{13}^{13}R_{16}^{T}(0),R_{17}^{T}(t)=A_{13}^{13}B_{11}^{11}R_{17}^{T}(0),\nonumber\\
      R_{18}^{T}(t)&=&A_{13}^{13}B_{12}^{12}R_{18}^{T}(0),R_{19}^{T}(t)=A_{13}^{13}B_{13}^{13}R_{19}^{T}(0),\nonumber\\
      R_{23}^{T}(t)&=&A_{11}^{11}B_{23}^{23}R_{23}^{T}(0),R_{24}^{T}(t)=A_{12}^{12}B_{21}^{21}R_{24}^{T}(0),\nonumber\\
      R_{25}^{T}(t)&=&A_{12}^{12}B_{22}^{22}R_{25}^{T}(0),R_{26}^{T}(t)=A_{12}^{12}B_{23}^{23}R_{26}^{T}(0),\nonumber\\
      R_{27}^{T}(t)&=&A_{13}^{13}B_{21}^{21}R_{27}^{T}(0),R_{28}^{T}(t)=A_{13}^{13}B_{22}^{22}R_{28}^{T}(0),\nonumber\\
      R_{29}^{T}(t)&=&A_{13}^{13}B_{23}^{23}R_{29}^{T}(0),\nonumber\\
      R_{34}^{T}(t)&=&A_{12}^{12}B_{31}^{31}R_{34}^{T}(0),R_{35}^{T}(t)=A_{12}^{12}B_{32}^{32}R_{35}^{T}(0),\nonumber\\
      R_{36}^{T}(t)&=&A_{12}^{12}(B_{33}^{11}R_{14}^{T}(0)+B_{33}^{22}R_{25}^{T}(0)+B_{33}^{33}R_{36}^{T}(0)),\nonumber\\
      R_{37}^{T}(t)&=&A_{13}^{13}B_{31}^{31}R_{37}^{T}(0),R_{38}^{T}(t)=A_{13}^{13}B_{32}^{32}R_{38}^{T}(0),\nonumber\\
      R_{39}^{T}(t)&=&A_{13}^{13}(B_{33}^{11}R_{17}^{T}(0)+B_{33}^{22}R_{28}^{T}(0)+B_{33}^{33}R_{39}^{T}(0)),\nonumber\\
      R_{45}^{T}(t)&=&A_{22}^{22}B_{12}^{12}R_{45}^{T}(0),R_{46}^{T}(t)=A_{22}^{22}B_{13}^{13}R_{46}^{T}(0),\nonumber\\
      R_{47}^{T}(t)&=&A_{23}^{23}B_{11}^{11}R_{47}^{T}(0),R_{48}^{T}(t)=A_{23}^{23}B_{12}^{12}R_{48}^{T}(0),\nonumber\\
      R_{49}^{T}(t)&=&A_{23}^{23}B_{13}^{13}R_{49}^{T}(0),\nonumber\\
      R_{56}^{T}(t)&=&A_{22}^{22}B_{23}^{23}R_{56}^{T}(0),R_{57}^{T}(t)=A_{23}^{23}B_{21}^{21}R_{57}^{T}(0),\nonumber\\
      R_{58}^{T}(t)&=&A_{23}^{23}B_{22}^{22}R_{58}^{T}(0), R_{59}^{T}(t)=A_{23}^{23}B_{23}^{23}R_{59}^{T}(0),\nonumber\\
      R_{67}^{T}(t)&=&A_{23}^{23}B_{31}^{31}R_{67}^{T}(0),R_{68}^{T}(t)=A_{23}^{23}B_{32}^{32}R_{68}^{T}(0),\nonumber\\
      R_{69}^{T}(t)&=&A_{23}^{23}(B_{33}^{11}R_{47}^{T}(0)+B_{33}^{22}R_{58}^{T}(0)+B_{33}^{33}R_{69}^{T}(0)),\nonumber\\
      R_{78}^{T}(t)&=&B_{12}^{12}(A_{33}^{11}R_{12}^{T}(0)+A_{33}^{22}R_{45}^{T}(0)+A_{33}^{33}R_{78}^{T}(0)),\nonumber\\
      R_{79}^{T}(t)&=&B_{13}^{13}(A_{33}^{11}R_{13}^{T}(0)+A_{33}^{22}R_{46}^{T}(0)+A_{33}^{33}R_{79}^{T}(0)),\nonumber\\
      R_{89}^{T}(t)&=&B_{23}^{23}(A_{33}^{11}R_{23}^{T}(0)+A_{33}^{22}R_{56}^{T}(0)+A_{33}^{33}R_{89}^{T}(0)),
\end{eqnarray}
and the other nondiagonal elements can be obtained by $R_{ij}^{T}(t)=(R_{ji}^{T}(t))^{\ast}$, where $R^{T}(t)$ is a Hermitian matrix. When the two atoms are identical, and likewise for the two cavities and for the two reservoirs, there will be $A_{ij}^{mn}=B_{i'j'}^{m'n'}$. In the following section, we only discuss this identical case.

\section{\fs Steady entanglement states of bipartite subsystems of the DJCM in non-Markovian environments}
In the standard basis $ \mathcal{B}^{T}=\{|11\rangle, |10\rangle, |01\rangle, |00\rangle, |ee\rangle, |eg\rangle, |ge \rangle, |gg \rangle\}$, we set that the initial state of the DJCM is
\begin{eqnarray}\label{EB15}
     \begin{array}{cccc}
      \rho(0)&=&(r|\phi^{+}\rangle\langle\phi^{+}|+\frac{1-r}{4}I)_{ab}\otimes|gg\rangle_{AB}\langle gg|\\
     \end{array},
\end{eqnarray}
where $|\phi^{+}\rangle=\frac{1}{\sqrt{2}}(|10\rangle+|01\rangle)$ is the Bell state, and $I$ is the $4\times4$ identity, $r(0\leq r\leq1)$ is the purity of the initial state, $A$ and $B$ indicate two atoms, $a$ and $b$ express two cavities. We can obtain the reduced density matrix $\rho_{ABab}$ at all times from Eq.~(\ref{EB13}) and Eq.~(\ref{EB14}). Then $\rho_{AB}$, $\rho_{ab}$, $\rho_{Aa}$, $\rho_{Bb}$, $\rho_{Ab}$ and $\rho_{aB}$ can also be obtained by taking a partial trace of $\rho_{ABab}$ over the other two degrees of freedom.

Particularly, we find that there are the steady entanglement states both in Markovian and in non-Markovian regimes. In the Markovian regime, bipartite subsystems can tend to steady states in a time if and only if the atom couples strongly to its cavity. Nevertheless, when the atom couples weakly to its cavity, bipartite subsystems can also tend to steady states in a time due to the memory and feedback effect of environment if and only if the non-Markovian effect is strong. It is very interesting that there are the same steady states in this two different conditions. When $\Omega\geq50\gamma_{0}$ and $2\gamma_{0}<\lambda\leq5\gamma_{0}$, or $\Omega\geq\gamma_{0}$ and $\lambda\leq0.1\gamma_{0}$, the steady entanglement states of bipartite subsystems can be respectively expressed as
\begin{eqnarray}\label{EB16}
\rho^{ss}_{Ab}=\rho^{ss}_{aB}=\rho^{ss}_{AB}=\rho^{ss}_{ab}=\left(
            \begin{array}{cccc}
              \frac{1-r}{64}&0&0&0\\
              0&\frac{7+r}{64}&\frac{r}{8}&0\\
              0&\frac{r}{8}&\frac{7+r}{64}&0\\
              0&0&0&\frac{49-r}{64}\\
            \end{array}
          \right)
\end{eqnarray}
and
\begin{eqnarray}\label{EB17}
\rho^{ss}_{Aa}=\rho^{ss}_{Bb}=\left(
            \begin{array}{cccc}
              0&0&0&0\\
              0&\frac{1}{8}&\frac{1}{8}&0\\
              0&\frac{1}{8}&\frac{1}{8}&0\\
              0&0&0&\frac{6}{8}\\
            \end{array}
          \right).
\end{eqnarray}

From Eq.~(\ref{EB16}) and Eq.~(\ref{EB17}), we know that the steady states $\rho^{ss}_{Ab}$, $\rho^{ss}_{aB}$, $\rho^{ss}_{AB}$ and $\rho^{ss}_{ab}$ are dependent on $r$, while the steady states $\rho^{ss}_{Aa}$ and $\rho^{ss}_{Bb}$ are $r$ independent.

\begin{center}
\includegraphics[width=16cm,height=10cm]{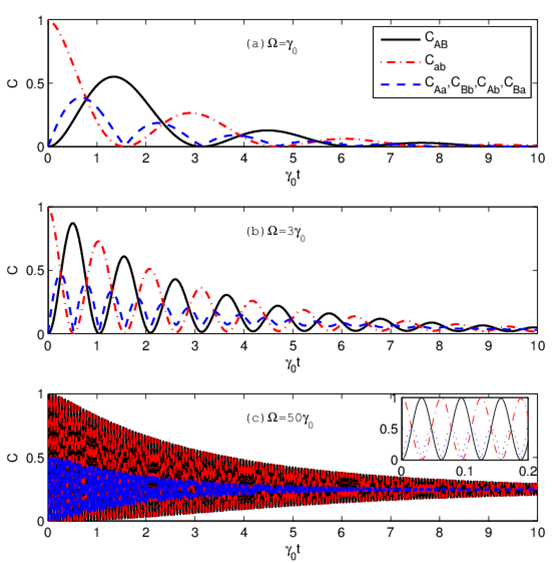}
\parbox{16cm}{\small{\bf Fig2.}
(Color online)The effect of the atom-cavity coupling $\Omega$ on the entanglements versus $\gamma_{0}t$ for $r=1$ in the Markovian regime($\lambda=5\gamma_{0}$). (a)$\Omega=\gamma_{0}$; (b)$\Omega=3\gamma_{0}$; (c)$\Omega=50\gamma_{0}$. $C_{AB}$(black, solid), $C_{ab}$(red, dotted-dashed), $C_{Aa}$($C_{Bb}$, $C_{Ab}$ and $C_{aB}$)(blue, dashed). The inset in Fig.2(c) shows the very short-time dynamics.}
\end{center}

\section{\fs Entanglement swapping of the DJCM in non-Markovian environments}
In order to quantify the entanglements of bipartite subsystems, we use Wootter's concurrence\cite{Bellomo1,Wootter}, which is defined as
\begin{equation}\label{EB119}
C=max(0,\sqrt{\lambda_{1}}-\sqrt{\lambda_{2}}-\sqrt{\lambda_{3}}-\sqrt{\lambda_{4}})
\end{equation}
where $\lambda_{i}$ are the eigenvalues, organized in a descending order, of the matrix $\tilde{\rho}
=\rho(\sigma_{y}\otimes\sigma_{y})\rho^{\ast}(\sigma_{y}\otimes\sigma_{y})$. And $C_{AB}$, $C_{ab}$, $C_{Aa}$, $C_{Bb}$, $C_{Ab}$ and $C_{aB}$ indicate the entanglements of the subsystems $AB$,  $ab$, $Aa$, $Bb$, $Ab$ and $aB$, respectively. In the following, we analysis the influences of the atom-cavity coupling, the non-Markovian effect and the initial state purity on the entanglement dynamics.

\begin{center}
\includegraphics[width=16cm,height=10cm]{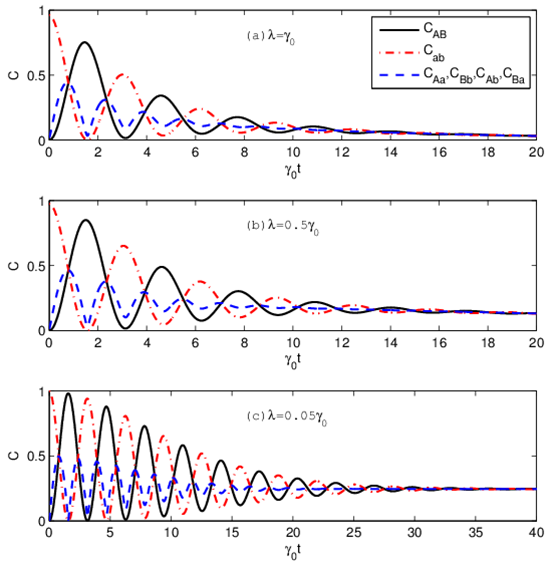}
\parbox{16cm}{\small{\bf Fig3.}
(Color online)The influence of the non-Markovian effect on the entanglements versus $\gamma_{0}t$ with $\Omega=\gamma_{0}$ and $r=1$. (a)$\lambda=\gamma_{0}$; (b)$\lambda=0.5\gamma_{0}$; (c)$\lambda=0.05\gamma_{0}$. $C_{AB}$(black, solid), $C_{ab}$(red, dotted-dashed), $C_{Aa}$($C_{Bb}$, $C_{Ab}$ and $C_{aB}$)(blue, dashed).}
\end{center}

In Fig.2, we plot the effect of the atom-cavity coupling $\Omega$ on the entanglement dynamics in the Markovian regime($\lambda=5\gamma_{0}$) with $r=1$. From Fig.2, we can see that the entanglements between any atom and any cavity is always equal when $r=1$, i.e. $C_{Aa}=C_{Bb}=C_{Ab}=C_{aB}$. In Fig.2(a), it can be found that, when $\Omega=\gamma_{0}$, as time $t$ increases, $C_{ab}$ will oscillate damply to zero from 1.0 under the cavity dissipation, while $C_{AB}$ increases to 0.55 from zero then oscillates damply to zero. This denotes that, the two atoms not to directly interact can be entangled though they are initially in a product state $|gg\rangle_{AB}$. Beside this, we also observe that, $C_{Aa}$($C_{Bb}$, $C_{Ab}$ and $C_{aB}$) rises to 0.49 from zero then oscillates damply to zero. That is, the entanglement between two cavities can be swapped to the two-atom subsystem and four atom-cavity subsystems via the interaction between the atom with its cavity. Comparing Fig.2(a) and Fig.2(b), we can see that, their entanglement dynamics is similar. The difference is in the oscillating frequency of entanglement and in the entanglement decay rate.
The oscillating frequency of entanglement in the latter case is about three times of the former. The entanglement decay rate is obvious smaller than in the first one. Hence, increasing $\Omega$, the oscillating frequency of entanglement will become quick and the entanglement decay will become slow. But in the Markovian regime, if $\Omega<50\gamma_{0}$, all entanglements will eventually decay to zero in a short time. While all entanglements will tend to 0.25 when $\Omega\geq50\gamma_{0}$, shown as Fig.2(c).

Fig.3 exhibits the influence of the non-Markovian effect on the entanglements with $\Omega=\gamma_{0}$ and $r=1$. Fig.3 shows that, in the non-Markovian regime, $C_{ab}$ can also be swapped to other bipartite subsystems and $C_{Aa}=C_{Bb}=C_{Ab}=C_{aB}$ when $r=1$. Comparing Fig.2(a) and Fig.3(a), we find out that the entanglement evolution in the non-Markovian regime reveals distinct features: although all entanglements will decay to zero in a time when $\lambda=\gamma_{0}$, the decay rate of entanglement in this case is obviously smaller than in the Markovian regime  due to the memory and feedback effect of environment. Comparing Fig.3(a) and Fig.3(b), it is seen that, the smaller the value of $\lambda$ is, the stronger the non-Markovian effect is, the slower the entanglements reduce. Particularly, all entanglements will tend to 0.25 in a time when the non-Markovian effect gets strong enough, Fig.3(c) exhibits the $\lambda=0.05\gamma_{0}$ case.

\begin{center}
\includegraphics[width=16cm,height=4.5cm]{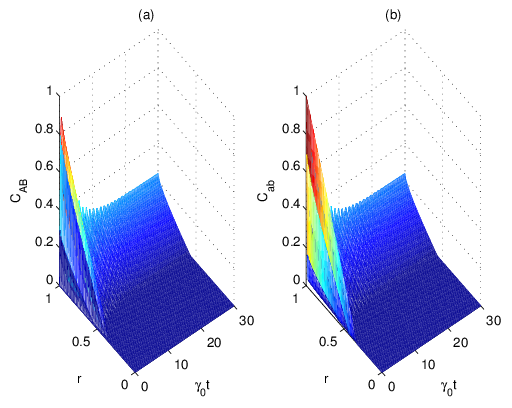}
\includegraphics[width=16cm,height=4.5cm]{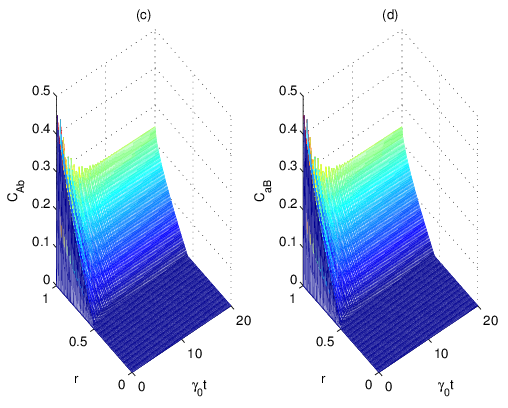}
\includegraphics[width=16cm,height=4.5cm]{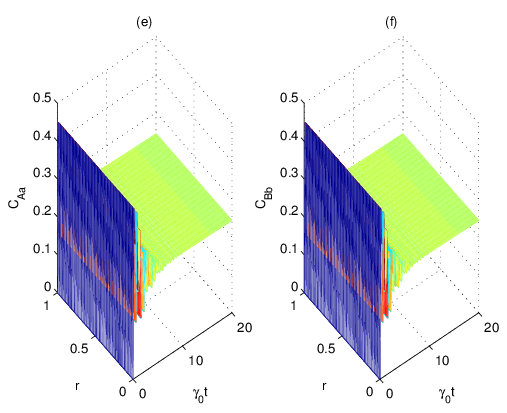}
\parbox{16cm}{\small{\bf Fig4.}
(Color online)The effect of the initial state purity $r$ on the entanglements versus $\gamma_{0}t$ with $\Omega=50\gamma_{0}$ and $\lambda=5\gamma_{0}$(the strong atom-cavity coupling and the Markovian regime). (a)$C_{AB}$; (b)$C_{ab}$; (c)$C_{Ab}$; (d)$C_{aB}$; (e)$C_{Aa}$; (f)$C_{Bb}$.}
\end{center}

Fig.4 displays the effect of the initial state purity $r$ on the entanglements with $\Omega=50\gamma_{0}$ and $\lambda=5\gamma_{0}$(i.e. the strong atom-cavity coupling and the Markovian regime). From Fig.4(a), we know that $C_{AB}$ is obvious $r$ dependent. For different $r$, the value of steady entanglement and the time it takes for $\rho_{AB}(t)\rightarrow\rho^{ss}_{AB}$ are different. The bigger the value of $r$ is, the larger the value of steady entanglement is. When $r>0.38$, $C_{AB}$ oscillates very quickly then decay to a steady value, but $C_{AB}$ is always equal to zero when $r<0.38$. Fig.4(b)-(d) plot the behaviors of $C_{ab}$, $C_{Ab}$ and $C_{aB}$ versus $\gamma_{0}t$, respectively. Comparing Fig.4(a) and Fig.4(b)-(d), we find that, the dynamics revolutions of $C_{AB}$, $C_{ab}$, $C_{Ab}$ and $C_{aB}$ are different before they get to the steady states, but they have the same steady value. Fig.4(e)-(f) give the relations of $C_{Aa}$ and $C_{Bb}$ versus $\gamma_{0}t$. It can be seen that $C_{Aa}$ is always equal to $C_{Bb}$ and both of them tend to 0.25 in a time, and their dynamics revolutions are all $r$ independent. It is worth noting that $C_{Aa}$ and $C_{Bb}$ can also oscillate damply to 0.25 even when $r=0$. The reason is that the atom $A(B)$ can be entangled with it cavity $a(b)$ through the coupling interaction of the atom-cavity when $\rho_{ab}(0)=\frac{1}{4}I$. Fig.5 shows the effect of the initial state purity $r$ on the entanglements versus $\gamma_{0}t$ with $\Omega=\gamma_{0}$ and $\lambda=0.05\gamma_{0}$(the weak atom-cavity coupling and the strong non-Markovian effect). In this case, the entanglement dynamics behaviors of six bipartite subsystems are similar respectively to those in Fig.4 except that the oscillating frequency in Fig.5 is much shorter than that in Fig.4. These results accord with the results from Eq.~(\ref{EB16}) and Eq.~(\ref{EB17}).

\begin{center}
\includegraphics[width=16cm,height=4.5cm]{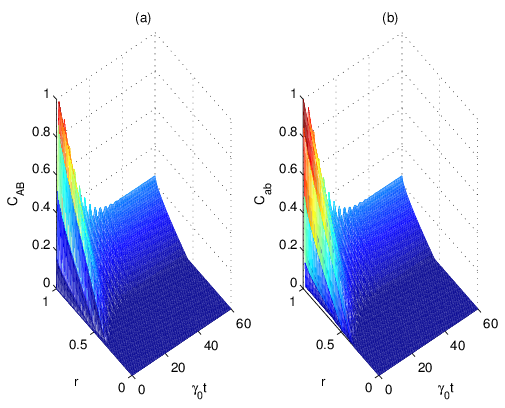}
\includegraphics[width=16cm,height=4.5cm]{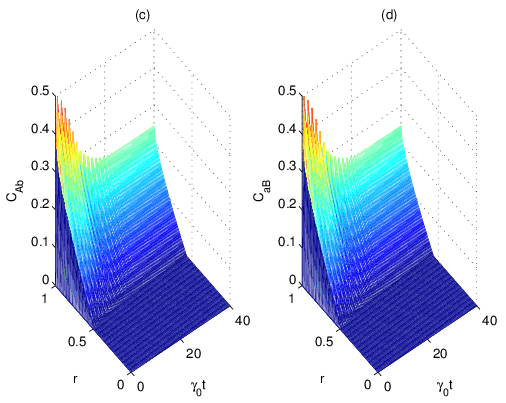}
\includegraphics[width=16cm,height=4.5cm]{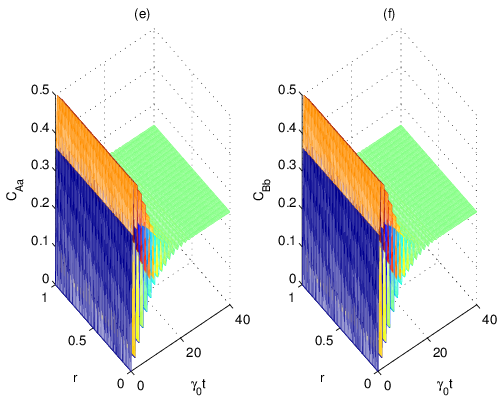}
\parbox{16cm}{\small{\bf Fig5.}
(Color online)The effect of the initial state purity $r$ on the entanglements versus $\gamma_{0}t$ with $\Omega=\gamma_{0}$ and $\lambda=0.05\gamma_{0}$(the weak atom-cavity coupling and the strong non-Markovian effect). (a)$C_{AB}$; (b)$C_{ab}$; (c)$C_{Ab}$; (d)$C_{aB}$; (e)$C_{Aa}$; (f)$C_{Bb}$.}
\end{center}

We may give the physical interpretations of the above results. In the Markovian regime and with small $\Omega$, the quantum information will continually dissipate to the environment during the two-cavity entanglement is swapped to other bipartite subsystems, so that all entanglements oscillate damply to zero, shown as Fig.2(a)-(b). However, when $\Omega$ is very large, the quantum entanglement can be exchanged very rapidly between bipartite subsystems so that the quantum information dissipated to the reservoir will obviously reduce. In a time, the entanglements can be effectively trapped in the DJCM, shown as Fig.2(c) and Fig.4. For the weak non-Markovian effect and with $\Omega=\gamma_{0}$, the entanglement decay rate will become small due to the memory and feedback effect of the non-Markovian environment but all entanglements will eventually decay to zero, shown as Fig.3(a)-(b). If the non-Markovian effect is very strong, the quantum information dissipated to the environment can be effectively fed back to the cavity so that the quantum entanglement can be effectively exchanged between bipartite subsystems, thus the entanglements will tend to a stationary value after a time, shown as Fig.3(c) and Fig.5.

\section{\fs Conclusion}
In conclusion, we have investigated the quantum entanglement dynamics of the DJCM interacting with external environments by the TCL master equation method. We obtain the analytical solution of this model and discuss in detail the influence of the atom-cavity coupling, the non-Markovian effect and the initial state purity on the entanglement dynamics when $\rho(0)=(r|\phi^{+}\rangle\langle\phi^{+}|+\frac{1-r}{4}I)_{ab}\otimes|gg\rangle_{AB}\langle gg|$. The results show that, the entanglement between two cavities can be swapped to other bipartite subsystems(i.e. $AB$, $Aa$, $Ab$, $aB$ and $Bb$) by interaction between the atom and its own cavity. We obtain the steady entanglement states of six bipartite subsystems, which provides a new method copying entanglement that six pairs of entanglement can be simultaneously prepared by a pair of entanglement in the non-Markovian environment. But, due to the dissipation of environments, the entanglements of six bipartite subsystems will eventually decay to zero when the atom couples weakly to its cavity and the non-Markovian effect is also weak. Bipartite subsystems can tend to steady entanglement states if and only if there is the strong atom-cavity coupling or the strong non-Markovian effect. The steady entanglement states $\rho_{Aa}^{SS}$ and $\rho_{Bb}^{SS}$ are independent on the purity while the steady states of other bipartite subsystems are dependent on the purity. These results may offer interesting perspectives for future applications of open quantum systems in quantum optical, microwave cavity QED implementations, quantum communication and quantum information processing.

The current experimental technologies\cite{Raimond} show that our proposals have a certain feasibility. For example, a circular Rydberg atom with the two circular levels with principal quantum numbers 51 and 50 which are called $|e\rangle$ and $|g\rangle$ respectively, the $|e\rangle\Leftrightarrow|g\rangle$ transition is at 51.1 GHz corresponding to the atomic decay rate $\gamma_{0}=33.3$Hz. In fact, the atom-cavity coupling $\Omega=50\gamma_{0}=1665$Hz is very small. That is, a slight coupling will bring on distinct results in the DJCM. The coupling that is required to prepare the steady entanglement state could be realized by Stark-shifting the frequency with a static electric field. The typical Stark shift is about 200kHz \cite{Hagley}, which is far more than 1665Hz. This shift is therefore large enough to prepare the steady entanglement state. Moreover, in cavity QED experiments, ultrahigh finesse Fabry-Perot super-conducting resonant cavities with quality factors $Q=4.2\times10^{10}$, corresponding to the spectral width $\lambda=7$Hz, have been realized\cite{Kuhr}. These values correspond to $\lambda/\gamma_{0}\approx0.2$ which represents a good non-Markovian regime.

\section{\fs Acknowledgments}
This work was supported by the Science and Technology Plan of Hunan Province, China(Grant no. 2010FJ3148) and the National Natural Science Foundation of China (Grant No.11374096).

\centerline{\hbox to 8cm{\hrulefill}}

\end{CJK*} 
\end{document}